\documentclass[12pt]{article}
\usepackage{graphicx}
\usepackage{graphics}
\usepackage{amssymb}
\usepackage{multirow}

\newcommand\pubnumber{SNSN-323-63}
\newcommand\pubdate{\today}

\def\bologna{INFN sez. di Bologna}

\def\Title#1{\begin{center} {\Large #1 } \end{center}}
\def\Author#1{\begin{center}{ \sc #1} \end{center}}
\def\Address#1{\begin{center}{ \it #1} \end{center}}

\newcommand\pubblock{\rightline{\begin{tabular}{l} \pubnumber\\
         \pubdate  \end{tabular}}}
\newenvironment{Abstract}{\begin{quotation}  }{\end{quotation}}
\newenvironment{Presented}{\begin{quotation} \begin{center} 
             PRESENTED AT\end{center}\bigskip 
      \begin{center}\begin{large}}{\end{large}\end{center} \end{quotation}}

\begin{document}
\begin{titlepage}
\pubblock

\vfill
\Title{Results from the OPERA experiment}
\vfill
\Author{Donato Di Ferdinando on behalf of OPERA Collaboration}
\Address{\bologna}
\vfill
\begin{Abstract}
The OPERA long baseline experiment located in the INFN Gran Sasso underground Laboratory was
designed to establish the $\nu_{\mu}$ $\rightarrow$ $\nu_{\tau}$ oscillations in appearance mode, using the CNGS
neutrino beam. Five $\nu_{\tau}$ candidate events have been detected in a data sample
from the 2008-2012 runs, with a expected background of 0.25 events. The background only hypothesis is rejected with a significance larger than 5${\sigma}$. The analysis of the tau neutrino sample in the
framework of the 3+1 neutrino model is presented. OPERA is able to identify $\nu_{e}$ CC events due to the good tracking capabilities of its target. The electron neutrino sample is also used to set limits on the 3+1 oscillation parameters. 

\end{Abstract}
\vfill
\begin{Presented}
NuPhys2015, Prospects in Neutrino Physics\\
Barbican Centre, London, UK,  December 16--18, 2015
\end{Presented}
\vfill
\end{titlepage}
\def\thefootnote{\fnsymbol{footnote}}
\setcounter{footnote}{0}

\section{The OPERA detector}
The OPERA (Oscillation Project with Emulsion tRacking Appararus) \cite{opera3} experiment was designed to conclusively prove the existence of $\nu_{\mu}$ $\rightarrow$ $\nu_{\tau}$ oscillations by the appearance of $\nu_{\tau}$ in a pure $\nu_{\mu}$ beam. The detector was located in the underground Gran Sasso Laboratory, 730 km from CERN, on the CNGS neutrino beam line. The $\nu_{\tau}$ signature is the detection of a  $\tau$-lepton (flight length of about 600 $\mu$m) produced by Charged Current (CC) interaction with the target mass of the detector. To cope with the challenging task of keeping a  high granularity in a large  detector mass the technique of Emulsion Cloud Chamber (ECC) has been used. The OPERA ECC target unit (called \textit{brick}) is composed of a sequence of 57 nuclear emulsion sheets 
interleaved with 56, $1~mm$ thick, lead plates for a mass of 8.3 kg; its transverse size is $12.5\times10.2~cm^2$ and the 
thickness along the beam direction is about 10 radiation lengths. A nuclear emulsion film is made of a 205 $\mu$m plastic base with a 44 $\mu$m sensitive layer on both sides. In the OPERA apparatus 150000  bricks were arranged in two target sections having each 28  vertical planes (called \textit{walls}) transverse to the beam; each wall was followed by two planes of plastic scintillator strips that composed the Target Tracker (TT) allowing to locate the brick in which the neutrino interaction occurred with a precision of $\mathcal{O}(cm)$. Each target section was completed by a magnetic spectrometer instrumented with Resistive Plate Chamber (RPC) detectors and high precision Drift Tubes (DT).

\section{Event reconstruction and analysis}
 The electronic detector data was used  to identify the brick in which the neutrino interaction occured. Events were classified as Charged Current like (1$\mu$) if a track was tagged as a muon or if a minimum number of detector walls was traversed, or as Neutral Current like (0$\mu$) in the complementary case \cite{mcs}.
The bricks tagged as containing a neutrino interaction vertex were extracted from the detector, their emulsion films were then  developed and sent to the scanning laboratories in Europe and Japan, where a search for $\tau$ decays was performed using custom automated optical microscopes.
The search for the neutrino interaction vertex starts from a set of track predictions, provided by the electronic detectors, found tracks are followed back in the brick, film by film,  from the most downstream one to the point where they originate. When a track disappears (i.e. is not found in three consecutive films) a volume of  1 cm$^{2}$ by 5 films upstream and 10 films downstream of the possible vertex point is scanned in order to fully explore the vertex area and possibly find other associated particle tracks. The search for the possible decay of a secondary track is then performed through a dedicated decay search procedure \cite{decayser}. Vertexes for which one track has an impact parameter larger than 10~$\mu$m are further analysed \cite{mcs,mcs0}, searching for secondary decay vertexes.
According to the decay topology of the event, either in one prong (electron, muon or hadron) or in three prongs, kinematical selection criteria are then applied to validate the candidate for appearance of the $\tau$ lepton.

\section{$\nu_{\mu}$ $\rightarrow$ $\nu_{\tau}$ oscillations search}
During the CNGS runs from 2008 to 2012, 19505 neutrino interactions were recorded in the target fiducial volume, corresponding to $17.97 \cdot 10^{19}$ protons on target (p.o.t.). The decay search procedure has been applied, after analysis of most of the data, to 5408 events. Out of this  sample, five $\tau$ candidates match the recontruction criteria  introduced  in the previous section. One of them is in the $\tau\rightarrow\mu$ decay channel \cite{tau3}, three in $\tau\rightarrow 1h$ \cite{tau1,tau4,tau5} and one in $\tau\rightarrow 3h$ \cite{tau2}.
In the analysed sample $0.25\pm0.05$ background events are expected, mainly from charmed events with an undetected primary muon, the remaining part coming from hadronic
re-interactions (for the hadronic decay channels) and large angle muon
scattering (for the $\tau \to \mu$ channel) as shown  in Tab.
\ref{tab:sigbkg}. Taking into account the different signal-to-noise ratio for
each decay channel, by the observation of five candidates a $5.1\sigma$
significance for the exclusion of the background-only hypothesis in the search for $\nu_\mu \to \nu_\tau$ oscillation is achieved \cite{tau5}.

\begin{table*}[htb]
\centering
  \resizebox{0.99\textwidth}{!}
            {%
              \begin{tabular}{c|cccc|c|c}
                \hline\hline
                \multirow{2 }{*}{Channel} &\multicolumn{4}{c|}{Expected background} & \multirow{2 }{*}{Expected signal} & \multirow{2}{*}{Observed} \\ \cline{2-5}
                &Charm &Had. re-interac. &Large $\mu$-scat. &Total & & \\\hline
                $\tau \rightarrow 1h$ 	&$0.017 \pm 0.003$  &$0.022 \pm 0.006$ 	&$-$  &$0.04 \pm 0.01$ &$0.52 \pm 0.10$ &3 \\ 
                $\tau \rightarrow 3h$ 	&$0.17 \pm 0.03$ 	&$0.003 \pm 0.001$ 	&$-$  &$0.17 \pm 0.03$ &$0.73 \pm 0.14$ &1\\ 
                $\tau \rightarrow \mu$ 	&$0.004 \pm 0.001$ 	&$-$ 	            &$0.0002 \pm 0.0001$  &$0.004 \pm 0.001$ &$0.61 \pm 0.12$ &1\\
                $\tau \rightarrow e$ 	&$0.03 \pm 0.01$ 	&$-$            	&$-$  &$0.03 \pm 0.01$ &$0.78 \pm 0.16$ &0\\
                \hline
                Total  &$0.22 \pm 0.04$ &$0.02 \pm 0.01$ &$0.0002 \pm 0.0001$ &$0.25 \pm 0.05$ &$2.64 \pm 0.53$ &5\\ \hline\hline
              \end{tabular} 

           }
            \caption{Expected signal and background events for the analysed data sample.}
            \label{tab:sigbkg}
\end{table*}

\section{Search for sterile neutrinos}

\begin{figure}[htb]
\centering
\includegraphics[height=2.5in]{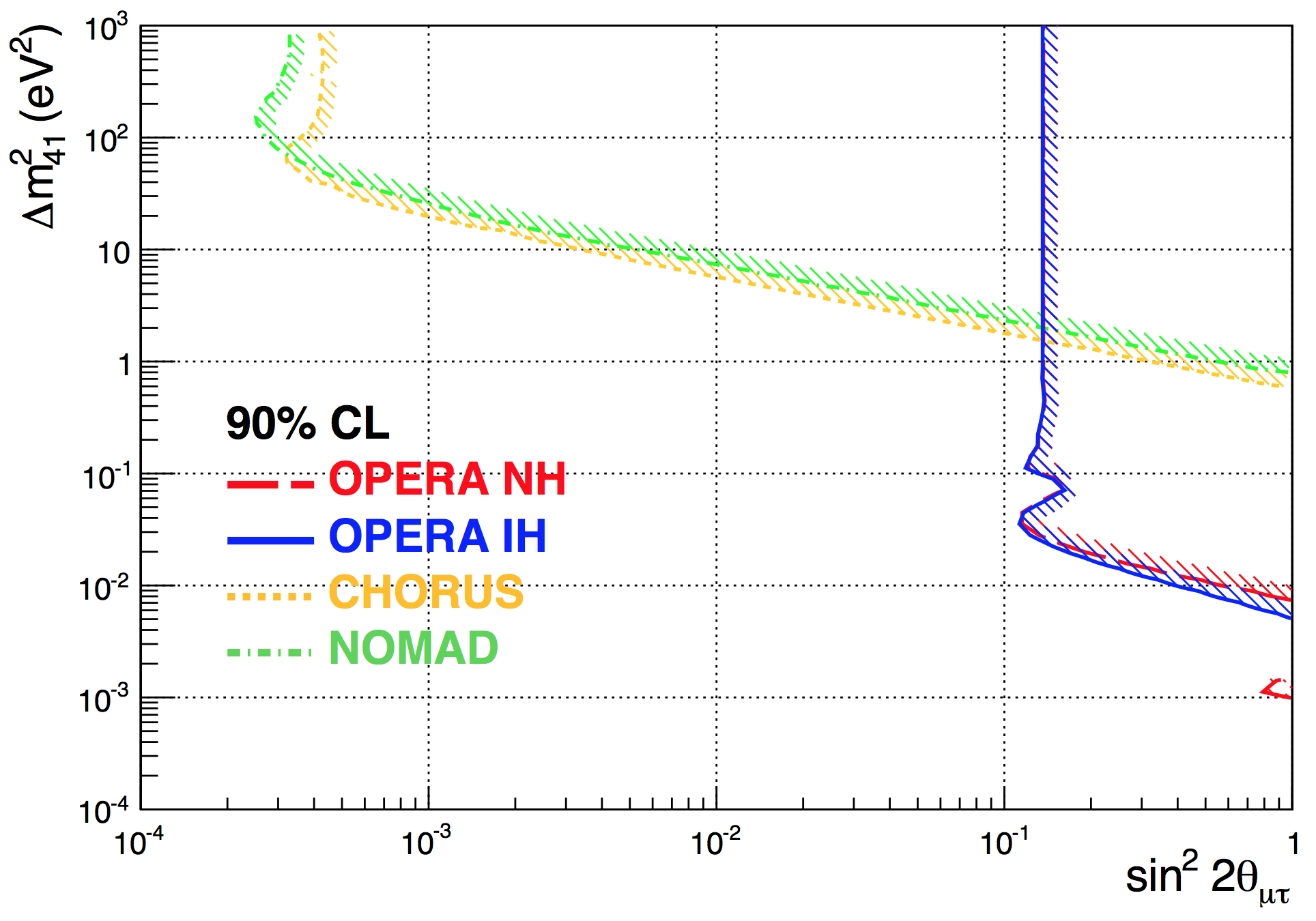}
\caption{\label{ster} OPERA $90\%$~C.L exclusion limits in the $\Delta m^2_{41}$  vs $\sin^{2}2\theta_{\mu\tau}$  parameter space for the normal (NH, dashed red) and inverted (IH, solid blue) hierarchy of the three standard neutrino masses. The exclusion plots by other experiments are also shown. Bands are drawn to indicate the excluded regions.
}
\end{figure}

The neutrino oscillation phenomenon is currently well described by the mixing of three neutrino flavor eigenstates ($\nu_{e}$, $\nu_{\mu}$ and $\nu_{\tau}$) with three mass eigenstates ($\nu_{1}$, $\nu_{2}$ and $\nu_{3}$). Nevertheless there are experimental results that cannot be accomodated within this framework \cite{anomalies}. 
These anomalies may hint to the existence of sterile neutrino(s) with $\Delta m^{2} \sim 1~\mbox{eV}^{2}$. The OPERA experiment can test the sterile neutrino hypothesis looking for deviations from predictions of the standard three-neutrino model.
OPERA $\nu_\tau$ appearance results have been used to derive limits on the mixing parameters of a massive ($\sim$ eV) sterile neutrino \cite{sterile}.
\\In presence of a fourth sterile neutrino with mass $m_4$, the oscillation probability is a function of the $4\times 4$ mixing matrix $U$ and of the three squared mass differences.
Observed neutrino oscillation anomalies, if interpreted in terms of one additional sterile neutrino, suggest $|\Delta m^2_{41}|$ values at the eV$^{2}$ scale. In the framework of the 3+1 model, at high values of $\Delta m^2_{41}$, the measured $90\%$~C.L. upper limit on the mixing term $\sin^{2}2\theta_{\mu\tau} = 4|U_{\mu4}|^{2}|U_{\tau4}|^{2}$ is 0.116, independently of the mass hierarchy of the three standard neutrinos. The OPERA experiment extends the exclusion limits on $\Delta m^{2}_{41}$ in the $\nu_{\mu}\rightarrow\nu_{\tau}$ appearance channel down to values of $10^{-2}$eV$^{2}$ at large mixing for $\sin^{2}2\theta_{\mu\tau}\gtrsim 0.5$ as shown in Figure \ref{ster}.

A systematic search for $\nu_e$ events was performed in the 2008-2009 data sample. Out of 505 0$\mu$ events with located interaction vertex, corresponding to an integrated intensity of $5.25\cdot10^{19}~$ p.o.t. \cite{nue}, 19 $\nu_{e}$ candidate events were observed. This number is compatible with the expected $\nu_e$ from the beam contamination ($19.8\pm2.8$).
The current result on the search for the three-flavour neutrino oscillation yields an upper limit $\sin^{2}2\theta_{13}< 0.44$ $(90\%$ C.L.).

OPERA limits the parameter space available for a non-standard $\nu_{e}$ appearance suggested by the results of the LSND and MiniBooNE experiments \cite{anomalies}. It further constrains the still allowed region around $\Delta m^2_{new} = 5\cdot10^{-2}$~eV$^{2}$. 
For large $\Delta m^2_{new}$ values, a Bayesian approach has been used and the upper limit on $\sin^2 2\theta_{new}$ reaches the value $7.2 \cdot 10^{-3}$.
\\These results will be updated using 50 $\nu_e$ candidates observed in the full data set.

\section{Conclusions}
The OPERA experiment taking data from 2008 to 2012, collected \nobreak{$17.97 \cdot 10^{19}$ p.o.t}.
\\Five $\nu_\tau$ candidates have been observed and the observation of $\nu_\mu \to \nu_\tau$ oscillations due to background only is  excluded at $5.1\sigma$.
\\The observed number of $\nu_e$ interactions is compatible with the non-oscillation hypothesis, but allows OPERA to set an upper limit in the parameter space for a non-standard $\nu_e$ appearance.
\\Limits on the mixing parameters of a massive sterile neutrino have also been derived  in the $\nu_{\mu}\rightarrow\nu_{\tau}$ appearance channel and the exclusion limits on $\Delta m^{2}_{41}$ have been extended down to values of $10^{-2}$ eV$^{2}$ at large mixing for $\sin^{2}2\theta_{\mu\tau}\gtrsim 0.5$.

\bibliographystyle{unsrt}

\end{document}